\documentclass[draftclsnofoot,onecolumn]{IEEEtran}
\usepackage{cite}

\usepackage{graphicx}
\usepackage{multicol}
\usepackage{epstopdf}
\usepackage{color}

\usepackage{algorithm}
\usepackage{algorithmic}
\usepackage{amsthm,amsmath,amsfonts}
\usepackage{bbm}
\usepackage{mathrsfs}

\usepackage{bm}
\usepackage{subfigure}

\begin{document}
%
\title{Power Control for a URLLC-enabled UAV system incorporated with DNN-Based Channel Estimation}
%
%
%
\author{Peng Yang, Xing Xi,
        Tony Q. S. Quek,~\IEEEmembership{Fellow,~IEEE},
        Xianbin Cao,~\IEEEmembership{Senior Member,~IEEE},
        Jingxuan Chen
\thanks{
P. Yang and T. Quek are with the Information Systems Technology and Design Pillar, Singapore University of Technology and Design, Singapore 487372.

X. Xi, X. Cao, and J. Chen are with School of Electronic and Information Engineering, Beihang University, Beijing, China. 
}
}

\maketitle

\begin{abstract}
This letter is concerned with power control for a ultra-reliable and low-latency communications (URLLC) enabled unmanned aerial vehicle (UAV) system incorporated with deep neural network (DNN) based channel estimation. Particularly, we formulate the power control problem for the UAV system as an optimization problem to accommodate the URLLC requirement of uplink control and non-payload signal delivery while ensuring the downlink high-speed payload transmission. This problem is challenging to be solved due to the requirement of analytically tractable channel models and the non-convex characteristic as well. To address the challenges, we propose a novel power control algorithm, which constructs analytically tractable channel models based on DNN estimation results and explores a semidefinite relaxation (SDR) scheme to tackle the non-convexity. Simulation results demonstrate the accuracy of the DNN estimation and verify the effectiveness of the proposed algorithm.

%
%
%

\end{abstract}

\begin{IEEEkeywords}
UAV communications, URLLC, Power control, DNN-based channel estimation
\end{IEEEkeywords}

%
\IEEEpeerreviewmaketitle

\section{Introduction}
%
%
%
%


\IEEEPARstart{R}{ecent} advances in manufacturing, communications, sensors, electronics, and control technologies have witnessed an unprecedented application increase of unmanned aerial vehicles (UAVs) in civilian and commercial domains, such as monitoring and detection, traffic control, emergency search and rescue, emergency communication recovery.
Among these applications, UAVs are responsible for transmitting high-speed endogenous data (sensed by onboard equipments) or exogenous data (received from other relays or base stations (BSs)) to a ground control station (GCS) or ground communication users. However, UAVs have stringent size, weight, and power constraints; therefore, energy-efficient data transmission should be closely considered in UAV communications.
Besides, the safety requirements are paramount. UAVs are obligated by the law to stay in the remote control range of a controller to ensure the safe flight. To this aim, control links for ultra-reliable and low-latency communications (URLLC) should be enabled \cite{ren2019achievable,ranjha2019quasi}.
In \cite{ren2019achievable}, the average achievable data rate for URLLC-enabled UAV systems was investigated, where a URLLC control link was established to deliver control information to a UAV for collision avoidance. The problem of how to optimize transmission distances and the packet blocklength in URLLC aided multi-hop UAV relay links for short packet transmission was studied in \cite{ranjha2019quasi}
Additionally, owing to service requirements among these applications, a UAV may fly in a three-dimensional (3D) trajectory, which brings great challenges for air-ground propagation channel modeling. In 3D space, the air-ground channel model depends not only on a UAV's horizontal location, but also its deployment altitude.
In \cite{al2017modeling}, the statistical behavior of the path-loss from a BS to a UAV was modelled.
Measurement-based modeling and analysis of UAV channels in the vertical dimension were performed in \cite{cui2019measurement}
Besides, the work in \cite{goudos2019artificial} explored artificial neural networks (ANNs) to generate numerical UAV channel coefficients. However, the channels obtained in \cite{al2017modeling,cui2019measurement,goudos2019artificial} are either statistical results that can only simulate channels in an average sense or numerical results without proposing theoretically analytical UAV channel models.

This letter investigates the power control for a URLLC-enabled UAV system incorporated with DNN-based channel estimation. The main contributions of this letter are as follows: 1) we formulate a BS and UAV transmit power control problem aiming at accommodating the URLLC requirement of delivering control and non-payload signal from the BS to the UAV while ensuring the high-speed payload transmission from the UAV to the BS; This problem, however, is highly challenging to be solved due to the lack of accurate and analytically tractable channel models, as well as the non-convex characteristic; 2) we construct accurate and analytically tractable channel models based on DNN estimation results; 3) with the DNN-based channel models, the power control problem is effectively solved after tackling the non-convex issue by exploring a semidefinite relaxation (SDR) scheme.

\section{system model and problem formulation}
This letter considers a uplink and downlink communication scenario. In this scenario, a 5G BS with a vertically placed $K$-element uniform linear array is utilized to transmit ultra-reliable and low-latency control (e.g., UAV flight trajectory) and non-payload (exactly, the UAV transmit power) signal via uplink wireless fading channels to control a single antenna UAV. The UAV is responsible for sending high-speed payload, i.e., video taken by the onboard camera, back to the 5G BS via downlink wireless fading channels. We focus on the communication problem of jointly optimizing the transmit power of the BS and the UAV to guarantee the high-speed downlink payload transmission while satisfying the ultra-reliable and low-latency requirement of the uplink control and non-payload signal delivery. To mathematically analyze this problem, we assume that the time is discretized into a sequence of time slots. The duration of a time slot is $T_f$ including a duration of $D_u$ for control signal delivery and a duration of $T_f - D_u$ for payload transmission. Denote $\bm x_{\rm B} = [x_{\rm B}, y_{\rm B}, g_{\rm B}]^{\rm T}$ as the coordinates of the BS and $\bm x_{\rm v}(t) = [x_{\rm v}(t), y_{\rm v}(t), g_{\rm v}(t)]^{\rm T}$ as the coordinate of the UAV at time slot $t$. We next model the BS-to-UAV (BtU) and the UAV-to-BS (UtB) channel.

\subsection{BtU and UtB channel models}
As some existing statistical channel models \cite{al2017modeling,cui2019measurement} are inaccurate when modelling channels of the local environment where UAVs are actually deployed.
To tackle this issue, we model the DNN-based BtU channel at time slot $t$ as follows:
\begin{equation}\label{eq:BtU_channel_gain}
h_{k}^{\rm B}(t) = G_{\rm B,v}(\bm x_{\rm B}, \bm x_{\rm v}(t)) G_r \bar g(\bm x_{\rm B}, \bm x_{\rm v}(t)) f_k^{\rm B}(t) \buildrel \Delta \over = \frac{\theta_{k}^{\rm ul}(t)}{{{D}_{{\rm B,v}}^2(t)}},
\end{equation}
where $h_{k}^{\rm B}(t)$ is the channel from the $k$-th antenna of the BS to the UAV, $G_{\rm B,v}(\bm x_{\rm B}, \bm x_{\rm v}(t))$ denotes the transmitting antenna gain, $G_r$ represents the receiving antenna gain, $\bar g(\bm x_{\rm B}, \bm x_{\rm v}(t))$ is the path loss between the BS and UAV, $f_{k}^{\rm B}(t)$ is a random small-scale fading. Besides, ${\theta_{k}^{\rm ul}(t)} := \mu(\bm s_k^{\rm ul}(t)|\theta^{\mu_k} (t))$ is a channel coefficient that will be determined based on the DNN described in subsection III-A. $\bm s_k^{\rm ul}(t)$ denotes the DNN input, and $\theta^{\mu_k} (t)$ is the DNN parameters. ${{D}_{{\rm B,v}}(t)}= {{||{{\bm x}_{\rm v}}(t) - {{\bm x}_{\rm B}}||_2}}$ is the distance between UAV and the BS at $t$.

Likewise, we model the DNN-based UtB channel at time slot $t$ as follows:
\begin{equation}\label{eq:UtB_channel_gain}
h_{k}^{\rm v}(t) = G_t G_r \bar g(\bm x_{\rm B}, \bm x_{\rm v}(t)) f_k^{\rm v}(t) \buildrel \Delta \over = \frac{\theta_{k}^{\rm dl}(t)}{{{D}_{\rm B, v}^2(t)}},
\end{equation}
where $h_{k}^{\rm v}(t)$ is the channel from the UAV to the $k$-th antenna of the BS, $G_t$ denotes the transmitting antenna gain of UAV, $f_k^{\rm v}(t)$ is a random small-scale fading.
$\theta_{k}^{\rm dl}(t) := Q(\bm s_k^{\rm dl}(t)|\theta^{Q_k}(t))$ is the channel coefficient that will be identified based on the DNN described in subsection III-B. $\bm s_k^{\rm dl}(t)$ is the DNN input, and $\theta^{Q_k}(t)$ is the DNN parameters.

The advantages of (\ref{eq:BtU_channel_gain}) and (\ref{eq:UtB_channel_gain}) are: the channel models can reflect the actual local environment parameters; the theoretical expressions of the channels are not complicated, which will make the communication problem analytically tractable.

\subsection{Transmission and resource constraints}
Next, we present the constraints of satisfying the ultra-reliable and low-latency requirement of control and non-payload signal delivery while guaranteeing the high-speed payload transmission.
To satisfy the low-latency requirement, control and non-payload packets are typically very short. Therefore, like \cite{Ren2020Joint}, we assume that the fading channel is a quasi-static Rayleigh fading channel over a time slot and changes independently and explore the data rate formula in finite blocklength regime to model the uplink achievable data rate.
Particularly, define the transmit beamformer of the BS towards the UAV at time slot $t$ as $\bm v(t) \in {\mathbb C}^{K}$ and $\bm V(t) = \bm v(t) \bm v^{\rm H}(t) \in {\mathbb R}^{K \times K}$. Let $\bm H_{\rm B,v}(t) = \bm h_{\rm B,v}(t) \bm h_{\rm B,v}^{\rm H}(t)$, where ${\bm h_{\rm B,v}(t) = [h_1^{\rm B}(t);\ldots;h_K^{\rm B}(t)]}$.
Since ${\rm tr}(\bm A \bm B) = {\rm tr}(\bm B \bm A)$ for matrices $\bm A$, $\bm B$ of compatible dimensions, we have $|\bm h_{\rm B,v}^{\rm H}(t)\bm v(t)|^2 =$ $(\bm h_{\rm B,v}^{\rm H}(t)\bm v(t))^{\rm H}\bm h_{\rm B,v}^{\rm H}(t)\bm v(t)=$ ${\rm tr}(\bm v^{\rm H}(t) \bm h_{\rm B,v}(t) \bm h_{\rm B,v}^{\rm H}(t) \bm v(t)) = $ ${\rm tr}(\bm H_{\rm B,v}(t) \bm V(t))$. Thus, given the transmission latency $\tau$ and the codeword decoding error probability $\varepsilon$, the approximately achievable data rate of the UAV can take the form \cite{Ren2020Joint}:
\begin{equation}\label{rate_urllc}
\begin{array}{l}
R^{\rm ul}(t; \tau) \approx W  {{\log }_2}( {1 + \frac{{{\rm tr}(\bm H_{\rm B,v}(t) \bm V(t))}}{{{N_0}W}}} ) - \\
\qquad \qquad \sqrt {\frac{{W{B_u}(t)}}{{\tau }}} {Q^{ - 1}}(\varepsilon ){{\log }_2}e,
\end{array}
\end{equation}
where $W$ denotes the system bandwidth, $N_0$ is the noise power spectral density, ${B}(t) = 1 - {\left( {1 + \frac{{{\rm tr}(\bm H_{\rm B,v}(t) \bm V(t))}}{{{N_0}W}}} \right)^{ - 2}}$ is the channel dispersion, $Q^{ - 1}\left(  \cdot  \right)$ is the inverse of $Q$-function.

Owing to the low-latency requirement, we assume that the delivery of a control and non-payload packet should be completed within the duration $D_u$. Let $F_u$ be the packet length, we then obtain the following transmission constraint:
\begin{equation}\label{eq:latency_constraint}
\frac{F_u}{R^{\rm ul}(t; \tau)} \le {D_u}.
\end{equation}

The signal-to-noise ratio (SNR) can reflect the quality of signal reception. A great SNR is essential for the correct demodulation of received signals. To this end, the SNR experienced by a receiver is recommended to be greater than 20 dB \cite{Teltonika}. With such a great SNR, $B(t)$ can be accurately approximated as one.

Different from the uplink control and non-payload signal delivery, we leverage Shannon formula to model the achievable data rate of the BS for downlink payload transmission. Specifically, define $\bm H_{\rm v,B}(t)= \bm h_{\rm v,B}(t) \bm h_{\rm v,B}^{\rm H}(t)$ and $\bm Z=\bm z \bm z^{\rm T}$, where $\bm h_{\rm v,B}^{\rm H}(t) = [h_1^{\rm v}(t);\ldots;h_K^{\rm v}(t)]$ and $\bm z \in {\mathbb R}^K$ is a column vector with each element being one. Then, we have $|\bm h_{\rm v,B}^{\rm H}(t)\bm z|^2=$ ${\rm tr}(\bm H_{\rm v,B}(t)\bm Z)$. Denote the achievable data rate of the BS by $R^{\rm dl}(t)$ at time slot $t$, $R^{\rm dl}(t)$ can thus be written as
\begin{equation}\label{eq:dl_rate}
R^{\rm dl}(t) = W{\log _2}\left( {1 + \frac{p(t){{\rm tr}(\bm H_{\rm v,B}(t)\bm Z)}}{{{N_0}W}}} \right)
\end{equation}
where $p(t)$ is the UAV transmit power at $t$.

To satisfy the high-speed transmission requirement, we assume that $R^{\rm dl}(t)$ should not be less than a data rate threshold (denoted by $R_e^{\rm th}$) at each time slot, i.e.,
\begin{equation}\label{eq:data_rate_constraint}
R^{\rm dl}(t) \ge R_e^{\rm th}
\end{equation}

Besides, as both the BS and the UAV have the maximum transmit power, we have
\begin{equation}\label{eq:BS_UAV_power_constraint}
{\rm tr }(\bm V(t)) \le p_{\rm B}^{\max }, \text{ } p(t) \le {p_{\rm v}^{\max }}.
\end{equation}
where $p_{\rm B}^{\max }$ and ${p^{\max }}$ are the maximum BS and UAV transmit power, respectively.

Our goal is to optimize the BS and UAV transmit power simultaneously to accommodate the ultra-reliable and low-latency requirement of the uplink control and non-payload signal delivery while ensuring the high-speed downlink payload transmission in an energy-efficient manner.
Combining with the above analysis, we can formulate the energy-efficient uplink and downlink communication problem at each time slot $t$ as follows:
\begin{subequations}\label{eq:original_problem}
\begin{alignat}{2}
& \mathop {{\rm{maximize  }}}\limits_{p(t),\bm V(t)} \text{ }  {E_{eu}}(t) = \frac{{{R^{\rm dl}}(t)}}{{R_e^{\max }(t)}} - \eta \left( {\frac{{p(t)}}{{{p_{\rm v}^{\max }}}} + \frac{{\rm tr}(\bm V(t))}{{p_{\rm B}^{\max }}}} \right)\\
& {\rm s.t: } \text{ } \bm V(t) = \bm v(t) \bm v^{\rm H}(t), \\
& (\ref{eq:latency_constraint}), (\ref{eq:data_rate_constraint}), (\ref{eq:BS_UAV_power_constraint}).
\end{alignat}
\end{subequations}
where $\eta$ represents an energy efficiency coefficient.

The solution of (\ref{eq:original_problem}) is challenging due to the unknown channels and the non-convexity incurred by the product of $\bm v(t)$. To tackle this challenge, we first design a DNN-based scheme to estimate the channels and then exploit a relaxation scheme to transform the non-convex problem into a convex one. The solution of (\ref{eq:original_problem}) is presented in the following section in detail.

\section{Problem solution}
\subsection{DNN-based channel coefficient estimation}
Owing to the flexibility of the deployment of UAV and the time-varying channel, the BtU and UtB channels depend on the local environmental parameters, the UAV deployment altitude and the UAV horizontal location in a rather sophistical way \cite{cui2019measurement,goudos2019artificial}. Besides, according to (\ref{eq:BtU_channel_gain}) and (\ref{eq:UtB_channel_gain}), the theoretical expressions of BtU and UtB channels can be achieved if the corresponding channel coefficients can be generated.
Therefore, DNNs, which are of practical usage for complicated function approximation, are exploited to generate the BtU and UtB channel coefficients.

We first describe the procedure of estimating BtU channel coefficients using DNNs. As $K$ channel coefficients should be estimated, the BS is enforced to construct and train $K$ BtU DNNs. For the $k$-th BtU DNN, it consists of the input space, network parameters, and output spaces, as presented below:

\textbf{Input space $\bm s_k^{\rm ul}(t)$:} At each time slot $t$, let $\bm s_k^{\rm ul}(t) = [{\bm x}_{\rm v}(t); {\bm x}_{\rm B}; k \bm z ]$. The column $k \bm z$ is involved in the input space to highlight the significance of the antenna index $k$; this strategy is named as \emph{dimension extension}. Besides, the BS and UAV locations are introduced in the input space because they are closely-related to the antenna gain.

\textbf{Network parameters $\theta^{\mu_k} (t)$:} We consider two fully-connected hidden layers. The \emph{ReLU} function is utilized as the activation function in the hidden layers. Besides, the network parameters are initialized by a Xavier initialization scheme. 

\textbf{Output space $o_k^{\rm ul}(t)$:} $o_k^{\rm ul}(t)$ is the estimated output value of the output layer. We set $\hat o_k^{\rm ul}(t) = \hat {\theta}_{k}^{\rm ul}(t)$ with $\hat {\theta}_{k}^{\rm ul}(t)$ being the uplink target channel coefficient. Many methods, e.g., the least square-based method \cite{qiu2018miso}, can be utilized to obtain the coefficient. In these methods, the UAV is responsible for feeding back the received signal strength to the BS. Besides, the \emph{linear} function is considered as the activation function in the output layer.

The experience replay technique is exploited to train each BtU DNN to break the correlation among training samples and avoid the DNN oscillation and divergence. The following Algorithm \ref{alg_BtU_channel_gain} summarizes the steps of training the DNN for BtU channel estimation.
\begin{algorithm}[!htp]
\caption{DNN-based BtU channel coefficient estimation}
\label{alg_BtU_channel_gain}
\begin{algorithmic}[1]
\STATE {\textbf {Initialize:}} DNN $\mu(\bm s_k^{\rm ul}(\hat t)|\theta^{\mu_k} (\hat t))$ with network parameters $\theta^{\mu_k} (\hat t)$. Replay buffer $R$ with capacity $C$ and minibatch size $|{{\mathcal T}_t}|$, training interval $T_{\rm tr}$.
\FOR{each episode $\hat t = 1, 2, \ldots$}
\STATE Calculate $\bm s_k^{\rm ul}(\hat t)$ based on the BS and UAV locations. The BS obtains $\hat o_k^{\rm ul}(\hat t)$ based on a least square-based method \cite{qiu2018miso}.
\STATE Store the transition $(\bm s_k^{\rm ul}(\hat t), \hat o_k^{\rm ul}(\hat t))$ in the buffer $R$.
\STATE If $\hat t$ is an integer multiple of $T_{\rm tr}$, then go to step 6.
\STATE If $\hat t \ge |{{\mathcal T}_t}|$, then sample a random minibatch of $|{{\mathcal T}_t}|$ transitions $(\bm s_k^{\rm ul}(m), \hat o_k^{\rm ul}(m))$ from $R$, and call the ADAM method to update ${\theta ^{{\mu _k}}(t)}$ by minimizing the loss function $L({\theta ^{{\mu _k}}(\hat t)})= \frac{1}{{|{{\mathcal T}_t}|}}\sum\nolimits_{\tau  \in {{\mathcal T}_t}} {{{\left( {{\hat {o}_k^{\rm ul}}(\tau) -  o_k^{\rm ul}(\tau)} \right)}^2}}$. 
\ENDFOR
\end{algorithmic}
\end{algorithm}

Meanwhile, the BS is imposed to construct and train other $K$ DNNs for UtB channel coefficient estimation. Likewise, for the $k$-th UtB DNN, it is composed of the input space, network parameters, and output spaces, as described below:

\textbf{Input space $\bm s_k^{{\rm dl}}(t)$:} The locations of BS and UAV and the dimension extension strategy are utilized to design the input space; that is, we let $\bm s_k^{{\rm dl}}(t) = [{\bm x}_{\rm B}; {\bm x}_{\rm v}(t); k\bm z]$.

\textbf{Network parameters $\theta^{Q_k} (t)$:} We consider two fully-connected hidden layers. The \emph{ReLU} is used as an activation function in hidden layers. The network parameters are initialized by a Xavier initialization scheme.

\textbf{Output space $o_{k}^{{\rm dl}}(t)$:} $o_{k}^{{\rm dl}}(t)$ is the output value of the output layer. We let the target output value $\hat o_{k}^{{\rm dl}}(t) = \hat \theta_{ij}(t)$ with $\hat \theta_{k}^{\rm dl}(t)$ being the downlink target channel coefficient. Some methods like the space-time signal subspace projection method \cite{cai2012simo} can be leveraged to achieve the single-input multiple-output channel coefficient.
The \emph{linear} activation function is selected in the output layer.

Since the main steps of training the DNN for UtB channel estimation are similar to Algorithm \ref{alg_BtU_channel_gain}, we omit them for brevity.

\subsection{Semidefinite relaxation}
Given the estimated channels, it is still difficult to solve (\ref{eq:original_problem}) due to its non-convexity. To tackle this issue, a SDR scheme is explored. Specifically, applying the property: $\bm V(t)=\bm v(t) \bm v^{\rm H}(t)$ $ \Leftrightarrow \bm V(t) \succeq 0$ and ${\rm rank}(\bm V(t)) \le 1$. According to the principle of the SDR scheme, the low-rank condition ${\rm rank}(\bm V(t)) \le 1$ can be directly dropped when solving (\ref{eq:original_problem}).
Besides, in (\ref{eq:original_problem}), $\tau$ is uncertain. Fortunately, it can be observed that we can increase $R^{\rm ul}(t; \tau)$ by scaling up $\bm V(t)$ and $\tau$ in the feasible region to enforce the constraint (\ref{eq:latency_constraint}). Moreover, increasing $\tau$ will not result in the decrease of the objective function (\ref{eq:original_problem}a); we, therefore, scale up $\tau$ to $D_u$. Via introducing an auxiliary variable $\varphi (t)$, (\ref{eq:original_problem}) can thus be reformulated as
\begin{subequations}\label{eq:transformed_problem}
\begin{alignat}{2}
& \mathop {{\rm{maximize  }}}\limits_{\bm V(t), p(t), \varphi (t)} \text{ }  E_{eu}(t)=\frac{\varphi (t)}{{R_e^{\max }(t)}} - \eta \left( {\frac{{p(t)}}{{{p_{\rm v}^{\max }}}} + \frac{{{\rm tr}(\bm V(t))}}{{p_{\rm B}^{\max }}}} \right)\\
& {\rm s.t: } \text{ } {{R^{\rm dl}}(t)} \ge \varphi (t), \text{ } \frac{F_u}{R^{\rm ul}(t; D_u)} \le {D_u}, \\
& {\rm tr}(\bm V(t)) \le p_{\rm B}^{\max}, \text{ } p(t) \le p_{\rm v}^{\max}, \\
& \bm V(t) \succeq 0, \varphi (t) \ge R_e^{\rm th}.
\end{alignat}
\end{subequations}

(\ref{eq:transformed_problem}) is now a convex problem that can be effectively solved by some optimization tools like MOSEK. It can be proved that the SDR for the transmit power $\bm V(t)$ is tight. Owing to the space limitation, we omit the proof and the author can find the similar proof in \cite{yang2020should}. Based on the tightness characteristic, we can perform the eigenvalue decomposition on the optimized $\bm V(t)$ to obtain the beamformer $\bm v(t)$ at each time slot $t$. In summary, the steps of solving (\ref{eq:transformed_problem}) are presented in Algorithm \ref{alg_joint_power_control}.
\begin{algorithm}[!htp]
\caption{Joint uplink and downlink power control}
\label{alg_joint_power_control}
\begin{algorithmic}[1]
\STATE {\textbf {Initialize:}} Run initialization steps of Algorithm \ref{alg_BtU_channel_gain}.
\FOR {each episode $\hat t = 1, 2, \ldots, 500$}
\STATE Steps 3-7 of Algorithm \ref{alg_BtU_channel_gain} to pre-train BtU DNNs.
\ENDFOR
\STATE Likewise, pre-train UtB DNNs for 500 episodes.
\FOR{each slot $t = 1, 2, \ldots, T$}
\STATE Observe the inputs $\bm s_k^{\rm ul}(t)$ and $\bm s_k^{\rm dl}(t)$, $\forall k$. Leverage pre-trained BtU and UtB DNNs to estimate $\theta_k^{\rm ul}(t)$ and $\theta_k^{\rm dl}(t)$, with which $h_k^{\rm B}(t)$ and $h_k^{\rm v}(t)$ are obtained using (\ref{eq:BtU_channel_gain}) and (\ref{eq:UtB_channel_gain}), respectively.
\STATE Optimize $\bm V(t)$ and $p(t)$ by solving (\ref{eq:transformed_problem}).
\STATE Steps 3-7 of Algorithm \ref{alg_BtU_channel_gain}. Likewise, train UtB DNNs.
\ENDFOR
\end{algorithmic}
\end{algorithm}
\begin{figure*}[!htb]
\centering
  \subfigure[MAPE of BtU DNNs with C$^2$T]{
    \label{fig:subfig:a} 
    \includegraphics[width=0.40\hsize]{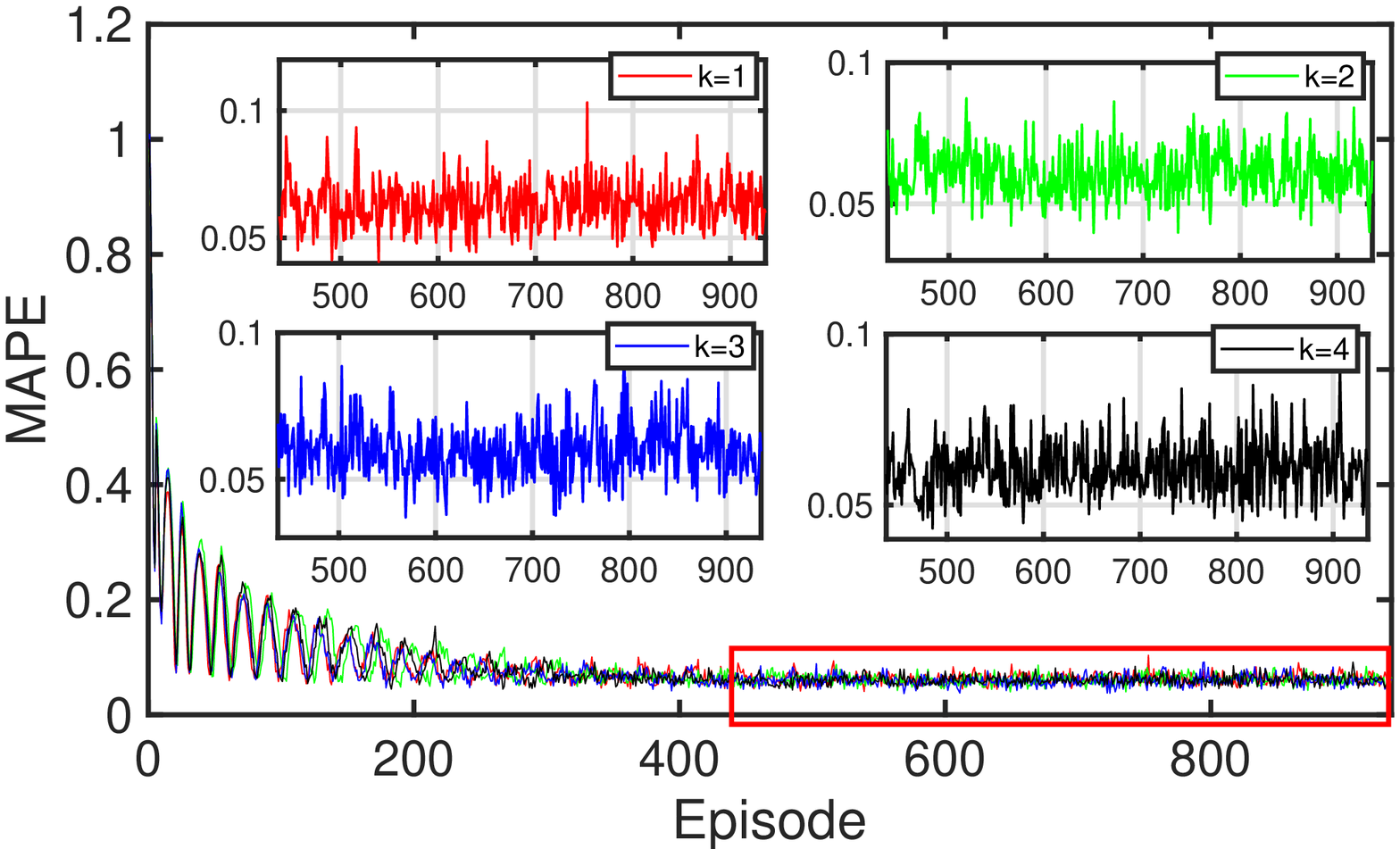}}  
  \subfigure[MAPE of BtU DNNs with VAT]{
    \label{fig:subfig:b} 
    \includegraphics[width=0.40\hsize]{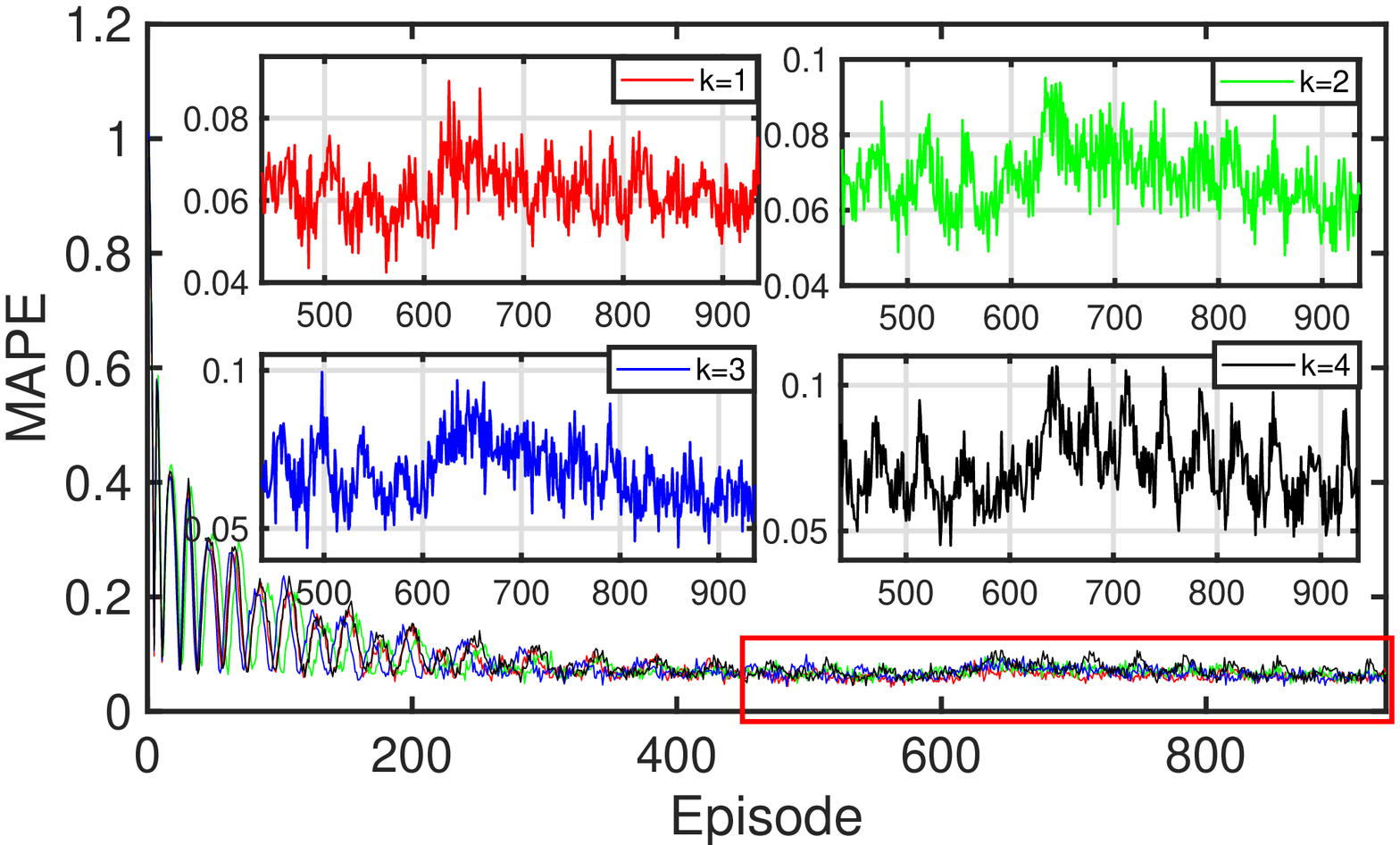}}
  \subfigure[MAPE of UtB DNNs with C$^2$T]{
    \label{fig:subfig:c} 
    \includegraphics[width=0.40\hsize]{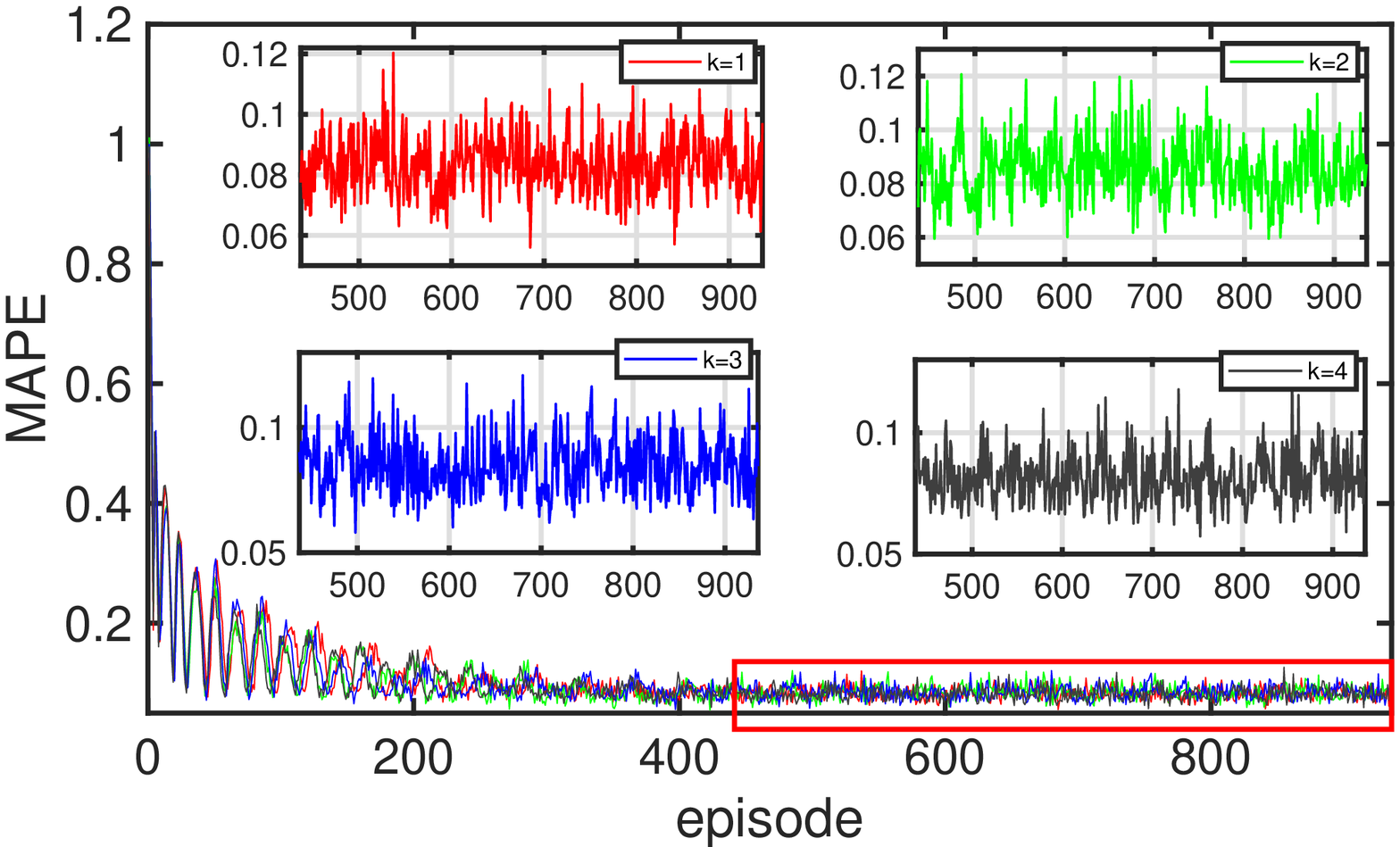}}
  \subfigure[MAPE of UtB DNNs with VAT]{
    \label{fig:subfig:d} 
    \includegraphics[width=0.40\hsize]{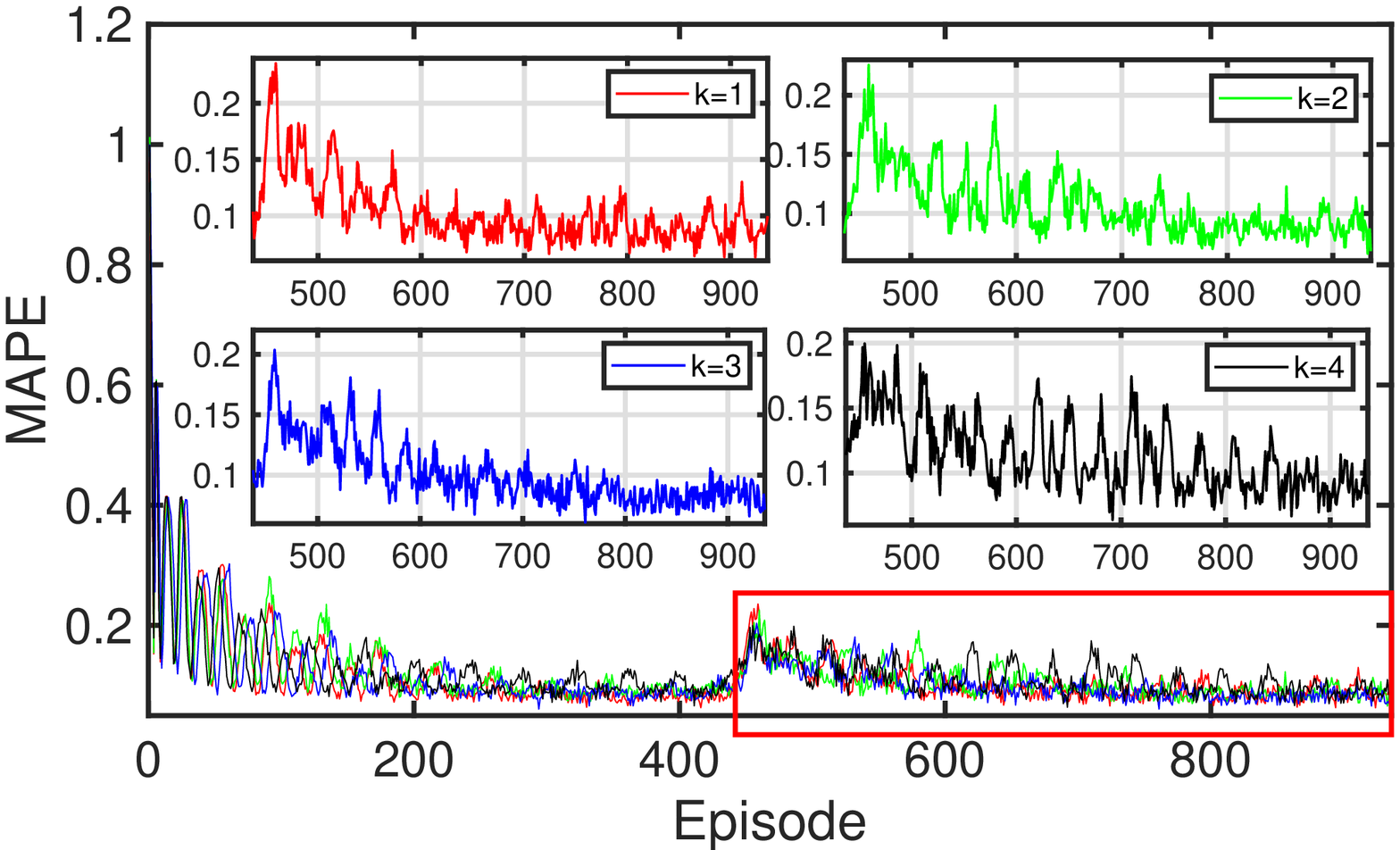}}
\caption{Accuracy of DNN-based channel estimation when the UAV follows a C$^2$T and a VAT.}
\label{fig_1}
\end{figure*}
\section{simulation results}
To verify the impact of the UAV horizontal location and UAV deployment altitude on the channel models, we design two types of UAV flying trajectories, i.e., circular trajectory (C$^2$T) and vertical ascent trajectory (VAT). The
Correspondingly, the circular trajectory-enabled power control algorithm (C$^2$T-PC) and vertical ascent trajectory-enabled power control algorithm (VAT-PC) are developed. In C$^2$T-PC, the UAV flies in a circular trajectory, and Algorithm \ref{alg_joint_power_control} is leveraged to perform the power control. In VAT-PC, the UAV flies in a vertical ascent trajectory, and the power control is conducted by running Algorithm \ref{alg_joint_power_control}.

In the simulation, we consider a $1$ km $\times$ $1$ km geographical area with high-rise buildings.
To accurately simulate BtU and UtB channels in a local environment, we generate building locations and heights based on a realization of a local building model suggested by International Telecommunication Union (ITU) $\rm P.1410-5$ with statistical parameters $\alpha = 0.3$, $\beta = 300$ buildings/km$^2$, $\gamma$ being modelled as a Rayleigh distribution with the mean value $\sigma = 30$ m. The heights of all buildings are clipped to not exceed 40 m for convenience.
With the building realization, the presence/absense of an LoS link between the BS and the UAV can be easily checked.
We next generate BtU and UtB path losses using the 3GPP TR $36.777$ model for urban Macro.
The small-scale fading coefficient is added assuming Rayleigh fading for the non line-of-sight (NLoS) case and Rician fading with 15 dB Rician factor for the line-of-sight (LoS) case.
Besides, the BS antenna model follows the 3GPP TR $36.873$ antenna model, where the antenna gain is related to the coordinates of the BS and the UAV.

The location of the center point, flying radius and flying altitude of the C$^2$T are $[0.5, 0.5, 0.05]^{\rm T}$ km, $0.375$ km, and $0.05$ km, respectively. For the VAT, the coordinates of the starting and ending points are $[0.5, 0.5, 0]^{\rm T}$ km and $[0.5, 0.5, 0.35]^{\rm T}$ km.
Other simulation parameters are as follows:
$\bm x_{\rm B} = [0.25, 0.375, 0.025]^{\rm T}$ km, $G_t=1$ dBi, $G_r=1$ dBi, $p_{\rm B}^{\max} = 5$ W, $p_{\rm v}^{\max} = 1$ W, ${\rm SNR}_{\rm th} = 20$ dB, $R_{e}^{\rm th} = 10$ Mb/s, $T_f = 5$ s, $N_0=-177$ dBm, $W = 20$ MHz, $F_u = 1000$ bits, $D_u = 1$ ms, $\varepsilon = 10^{-7}$. 
\begin{figure}[!htb]
\centering
  \subfigure[${\rm tr}(\bm V(t))$ and $p(t)$ vs. time slot $t$]{
    \label{fig:bs_subfig:a} 
    \includegraphics[width=2.6in, height=1.38in]{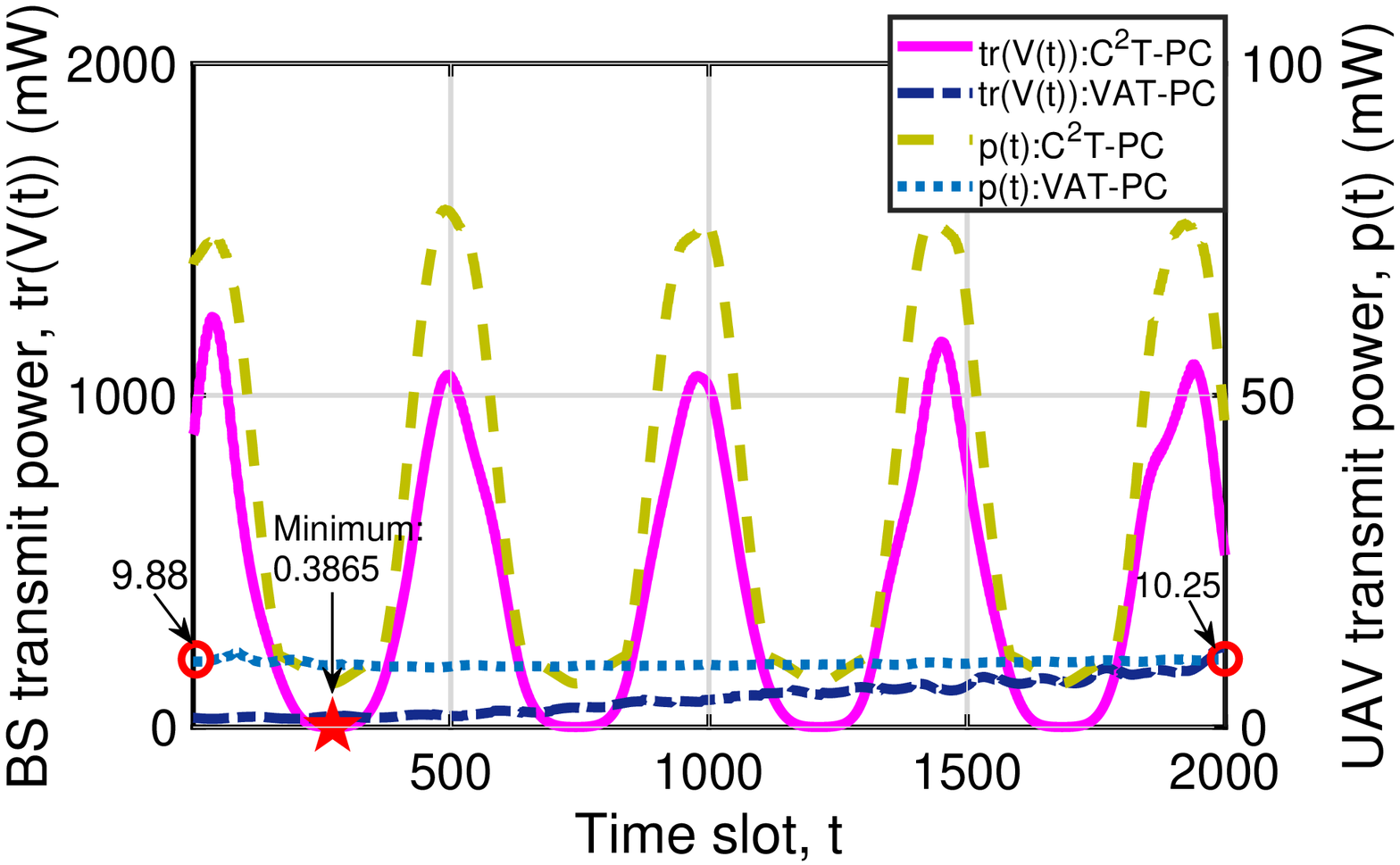}} \\  
  \subfigure[$R_e(t)$ and $E_{eu}(t)$ vs. time slot $t$]{
    \label{fig:rate_subfig:b} 
    \includegraphics[width=2.6in, height=1.38in]{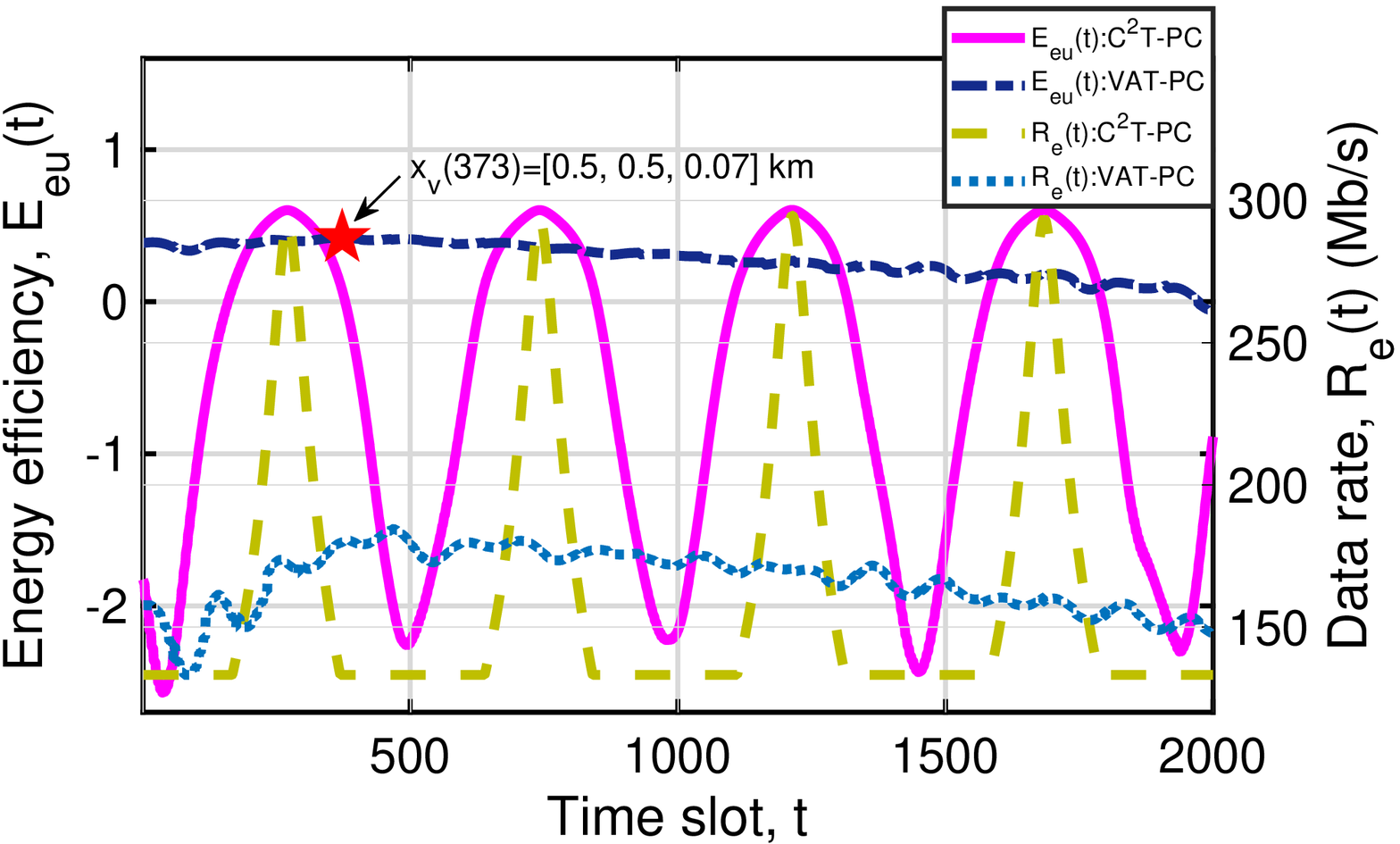}}
\caption{Trends of performance evaluation indexes vs. time slot $t$.}
\label{fig_performance}
\end{figure}

To testify the accuracy of the channel coefficient estimation algorithm, we plot the tendency of the mean absolute percent error (MAPE) \cite{goudos2019artificial} between the estimated channel coefficients and their target coefficients of all BtU and UtB DNNs in Fig. \ref{fig_1}. For the $k$-th BtU DNN, its MAPE is calculated by $MAPE= \frac{1}{{|{{\mathcal T}_t}|}}\sum\nolimits_{\tau  \in {{\mathcal T}_t}} {{{| ({{\hat {o}_k^{\rm ul}}(\hat t) -  o_k^{\rm ul}(\tau)})/{{\hat {o}_k^{\rm ul}}(\tau)} |}}}$ \cite{goudos2019artificial}. Besides, in Fig. \ref{fig_1}, the obtained MAPE values in the first 436 episodes and the later 500 episodes show the estimation accuracy of all BtU and UtB DNNs during the pre-training phase and the online training phase, respectively.

The following observations can be obtained from this figure: 1) all DNNs can converge after a period of training. Particularly, estimation errors of all DNNs are great initially, but tend to small values as more experience is accumulated. For instance, the MAPE values of all DNNs decrease from almost one to 0.2 after 200 episodes; 2) estimation errors of all DNNs are close to or less than 0.1 when the training phase is terminated; 3) in Fig. \ref{fig:subfig:d}, it is interesting to find that the MAPE value increases when the online train is activated. Yet, the MAPE value decreases as more fresh experience can be explored to train UtB DNNs. Note that the UAV is flying aimless during the pre-training phase while following the ascending trajectory during the online training phase for VAT-PC; thus, this result reflects the complexity of UtB channel modelling and the necessity of designing DNN-based UAV channels; 4) both UAV horizontal locations and UAV deployment altitude will significantly affect UAV channels; 5) Algorithm \ref{alg_BtU_channel_gain} can achieve good estimation results although actual BtU and UtB channel coefficients change rapidly with the flight of the UAV.

To verify the performance of C$^2$T-PC and VAT-PC, we plot the tendency of ${\rm tr}(\rm V(t))$, $p(t)$, $R_e(t)$, and $E_{eu}(t)$ in Fig. \ref{fig_performance}. From this figure, we observe that: 1) for C$^2$T-PC, its achieved BS and UAV transmit power ${\rm tr}({\bm V}(t))$, $p(t)$, energy efficiency $E_{eu}(t)$, and downlink achievable data rate $R_e(t)$ approximately periodically varies with $t$ due to the circular UAV trajectory; 2) for VAT-PC, its achieved ${\rm tr}({\rm V}(t))$ and $p(t)$ increase as the UAV rises owing to an increasing BtU distance; 3) for VAT-PC, the maximum $E_{eu}(t)$ is neither achieved at the UAV starting position nor the end position, but at a position between the starting position and the end position. This result also applies to the maximum $R_e(t)$. Besides, this result verifies the complicated impact of the UAV deployment altitude on the algorithm performance.

\section{Conclusion}
This letter investigated the power control for a URLLC-enabled UAV system incorporated with DNN-based channel estimation.
A non-convex power control problem for the UAV system was formulated to satisfy the URLLC requirement of uplink control and non-payload signal delivery while ensuring the downlink high-speed payload transmission.
To solve this problem, we first constructed analytically tractable channel models based on DNN estimation results. With the learned results, a joint uplink and downlink power control algorithm was then proposed after addressing the non-convexity via semidefinite relaxation.

\ifCLASSOPTIONcaptionsoff
  \newpage
\fi



\bibliographystyle{IEEEtran}
%
\bibliography{UAV_Channel_Gain_WCL}

\end{document}